\documentstyle[twoside,fleqn,espcrc2,epsfig,amssymb]{article}
\def\ct#1{{\cal #1}}
\newcommand{\lqcd}{\Lambda_{ QCD}}
\newcommand{\be}{\begin{equation}}
\newcommand{\eea}{\end{eqnarray}}
\newcommand{\beq}{\begin{equation}}
\newcommand{\eeq}{\end{equation}}
\newcommand{\bea}{\begin{eqnarray}}
\newcommand{\ee}{\end{equation}}

\newcommand{\dl}{\stackrel{\leftarrow}{D}}
\newcommand{\dr}{\stackrel{\rightarrow}{D}}
\newcommand{\dlr}{\stackrel{\leftrightarrow}{D}}
\newcommand{\db}{\bar{d}}
\newcommand{\gd}{\gamma_\mu}
\newcommand{\gu}{\gamma^\mu}

\newcommand{\nn}{\nonumber}

\newcommand{\Pj}{\mbox{I}\!\!\mbox{P}}
\newcommand{\Pjpv}{\mbox{1}\!\!\mbox{$^\wp$}}
\newcommand{\ctLambda}{\mbox{$\cal{J}$}\!\!\mbox{$\setminus$}}
\newcommand{\id}{\mbox{1$\!\!$I}}

\newcommand{\<}{\langle}
\renewcommand{\>}{\rangle}
\newcommand{\Tr}{\mbox{Tr}\;}

\title{HADRONIC WEAK INTERACTIONS OF LIGHT QUARKS}
\author{G.~Martinelli
\address{Dip. di Fisica, Univ. ``La Sapienza" and INFN, P.le A. Moro, I-00185 Rome, Italy}}       
\begin{document}
\begin{abstract}
In this review, three subjects are discussed:
a phenomenological application of lattice predictions 
 to $K^0$--$\bar K^0$ mixing in Super Symmetry; a discussion of the 
non-perturbative renormalization methods for four-fermion
operators and a new approach to extract weak matrix elements in effective 
theories  denoted as OPE without OPE (operator product expansion
without operator product expansion).\end{abstract}
\maketitle
\section{Introduction}
\label{sec:intro}
Since  the original proposals of using lattice QCD to study hadronic 
weak decays~\cite{CMP}--\cite{bernardargo}, 
substantial theoretical and experimental progress has been made:
with both Wilson and staggered  Fermions,
the main theoretical aspects of the renormalization of composite four-fermion
operators  are, by now, fully 
understood~\cite{boc,shape1}; the calculation of the $K^0$--$\bar K^0$ 
mixing amplitude, 
expressed in terms of the so-called renormalization group
invariant $B$-parameter $\hat B_K$, has reached a level of
precision (at least in the quenched approximation)  which is unpaired by any
other approach~\cite{ksgp}--\cite{jap1} (for a detailed discussion
see~\cite{Sharpe96}); 
increasing precision has also been gained in the determination of the
electro-weak penguin amplitudes (and of the strange quark mass~\cite{strange}),
which are relevant in the prediction
of the CP violating parameter $\epsilon^\prime/\epsilon$.
A lot of progress  has also been made in the determination of quantities which 
enter heavy hadron weak processes and which are reviewed  by
T.~Draper in his plenary talk~\cite{draper}.
Still, in particular for kaon physics, many problems are  unsolved
or poorly understood.  By using chiral perturbation theory and $SU(3)$
symmetry, for example, one may relate  $\hat B_K$ to the physical $\Delta I=3/2$
$K \to \pi \pi$ amplitude, $A_2$.  The large value of $\hat B_K \sim 0.85$ 
found in lattice calculations leads
to a prediction for $A_2$ which is about $40 \%$ larger than its experimental
value. It is not clear whether this is due to a failure of
chiral perturbation theory, or to electro-magnetic effects~\cite{don}. In the 
former case, the use of $K \to \pi$ matrix elements,
combined with chiral perturbation theory,  to predict the  $\Delta I=1/2$
amplitude $A_0$ or 
$\epsilon^\prime/\epsilon$ would  be very suspicious. Until very recently,
the only  existing calculation
of two other $B$-parameters which are very important for 
$\epsilon^\prime/\epsilon$, namely $B_5$ and $B_6$, was the one performed  
 with staggered Fermions
in~\cite{sharpe2}. 
Then,  the perturbative one-loop lattice corrections 
to the relevant  operators were computed~\cite{spis} and found  so large 
as to make the old estimate of \cite{sharpe2} very doubtful (a recent calculation
of $B_6$ was presented in~\cite{kkk}). Thus, we do not have any  reliable
lattice  estimate of $B_5$ and $B_6$.
Finally, there is no convincing lattice  calculation of $A_0$ yet, although some new
results with staggered fermions have been presented at this  Conference~\cite{pek}. 
\par In spite of considerable technical improvements, and of the
increasing computer power, no fundamental advance has been done 
last  years for the problems mentioned above. For this reason, 
 rather than presenting new results, I  discuss new  methods
which have been proposed in order to compute the relevant matrix elements.
In  most of the cases, these  methods have not been applied yet, or only
feasibility studies  exist.  This is the content of 
 sec.~\ref{sec:opewope}.  In sec.~\ref{sec:susy}, I  present
a recent  application of 
lattice calculations of $\Delta S=2$ amplitudes to flavor changing
neutral current (FCNC) phenomenology
in Super Symmetric models. In sec.~\ref{sec:npm}
several approaches, which have been used or proposed
for the non-perturbative renormalization of
lattice four-fermion operators, are discussed and compared.   
\section{$\Delta S=2$ transitions in SUSY}
\label{sec:susy}
The prescription of minimality in the number of new particles
introduced to supersymmetrise the Standard Model (SM), together with
the demand of conservation of baryon and lepton numbers, does not
prevent the appearance of more than 100 new SUSY parameters, in
addition to the 18 already present in the SM.
FCNC and CP violating phenomena are  protagonists of a drastic
reduction of these extra   degrees of freedom. Among the possible FCNC 
processes,  $K^0$--$\bar K^0$ ($B^0$--$\bar B^0$) mixing play a very special 
role.  In this section,  a recent analysis of $K^0$--$\bar K^0$ 
mixing in SUSY is discussed~\cite{silve}. This analysis makes use of    the recently computed NLO 
corrections to  the SUSY effective Hamiltonian~\cite{nlosusy} and  of
the matrix elements of the  relevant operators (renormalized 
non-perturbatively in the RI-MOM scheme), computed on the lattice~\cite{dosusy}.  
This is a new, interesting application of the lattice method for studying 
the physics beyond the SM.  Unfortunately,  for lack of space, many other applications 
to physics beyond the SM cannot be  discussed here. 
A detailed discussion of the non-perturbative methods which can be used to 
renormalize the relevant four-fermion operators can be found in 
sec.~\ref{sec:npm}. 
\par
\begin{figure}   % produce figure here
%    \centering
    \epsfxsize=0.50\textwidth
    \leavevmode\epsffile{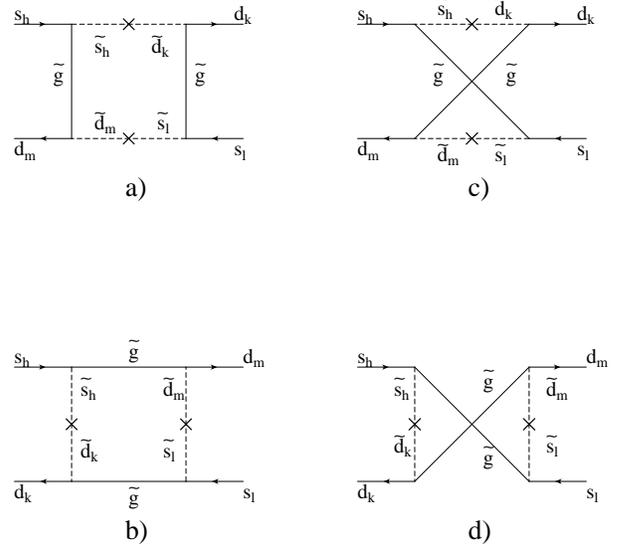}
\caption{Feynman diagrams for $\Delta S=2$ transitions, with
  $h,k,l,m=\{L,R\}$.}  
\label{fig:ds2}
\end{figure}

 Let us  start by defining the model used to construct the low-energy 
effective Hamiltonian for $\Delta F=2$ transitions.
 In the so-called mass insertion
approximation \cite{hall},  one chooses the super-CKM basis for the
fermion and sfermion states, where all the couplings of these particles
to neutral gauginos are flavour diagonal, while the genuine SUSY FC
effects  are  exhibited by the non-diagonality of the sfermion mass
matrices. Denoting by $\Delta^2$ the off-diagonal terms in the sfermion
mass matrices (i.e.  the mass terms relating sfermions of the same
electric charge, but different flavour), the sfermion propagators can
be expanded as a series in terms of $\delta = \Delta^2/ \tilde{m}^2$,
where $\tilde{m}$ is the average sfermion mass.  As long as $\Delta^2$
is significantly smaller than $\tilde{m}^2$, we can just take the
first term of this expansion and, then, the experimental information
concerning FCNC and CP violating phenomena translates into upper
bounds on the $\delta$s \cite{gabbiani}--\cite{GGMS}. 
There exist four different $\Delta$ mass-insertions connecting
flavours $d$ and $s$ along a sfermion propagator:
$\left(\Delta^d_{12}\right)_{LL}$, $\left(\Delta^d_{12}\right)_{RR}$,
$\left(\Delta^d_{12}\right)_{LR}$ and
$\left(\Delta^d_{12}\right)_{RL}$.  The indices $L$ and $R$ refer to
the helicity of the fermion partners.  For gluino-mediated
processes, for example, the amplitude for $\Delta S=2$
transitions in the full theory   is given by the computation
of the diagrams in fig.~\ref{fig:ds2}.
%___________________________________________________________________________
%___________________________________________________________________________
Having calculated the amplitude from the diagrams in
fig.~\ref{fig:ds2}, one has to choose a basis of local operators and
perform the matching of the full theory to the one described by ${\cal
  H}_{\rm eff}^{\Delta S=2}$. In~\cite{silve,dosusy}, they have adopted the form 
\begin{equation}
  \label{eq:defheff}
  {\cal H}_{\rm eff}^{\Delta S=2}=\sum_{i=1}^{5} C_i\, Q_i +
  \sum_{i=1}^{3} \tilde{C}_i\, \tilde{Q}_i
\end{equation} 
where 
\begin{eqnarray}
        Q_1 & = & \db^{\alpha}_L \gd s^{\alpha}_L \db^{\beta}_{L} \gu
        s^{\beta}_L\; , 
        \quad 
        Q_2  =  \db^{\alpha}_R  s^{\alpha}_L \db^{\beta}_R s^{\beta}_L\; ,
        \nn \\
        Q_3 & = & \db^{\alpha}_R  s^{\beta}_L \db^{\beta}_R s^{\alpha}_L\; ,
        \label{Qi}  \\
        Q_4 & = & \db^{\alpha}_R  s^{\alpha}_L \db^{\beta}_L s^{\beta}_R\; ,
        \quad
        Q_5  =  \db^{\alpha}_R  s^{\beta}_L \db^{\beta}_L s^{\alpha}_R\; ,
        \nn
\end{eqnarray}
and the operators $\tilde{Q}_{1,2,3}$ are obtained from the
$Q_{1,2,3}$ by the exchange $ L \leftrightarrow R$.  Here
$q_{R,L}=P_{R,L}\,q$, with $P_{R,L}=(1 \pm \gamma_5)/2$, and $\alpha$
and $\beta$ are colour indices.
The Wilson coefficients $C_i(\mu)$ ($\tilde{C}_i(\mu)$) for the operators 
renormalized at the scale $\mu$ are obtained by computing the coefficients 
at the large energy scale, $C_i(M_S)$, where $M_S$ is the order of the 
gluino (squark) mass, and then evolving them to $\mu$ using  suitable  QCD
renormalization group equations, which depend on the operator 
anomalous-dimension matrix. The latter is only determined from QCD and does not know
about the underlying fundamental theory at the large energy scale. 
The next-to-leading anomalous-dimension matrix for the most general
${\cal H}_{\rm eff}^{\Delta F=2}$ has been  computed in~\cite{nlosusy}. 
 They used the regularisation-independent (RI) anomalous
dimension in the Landau gauge, since we will make use of matrix
elements computed in lattice QCD with the same choice of
renormalisation scheme~\cite{dosusy}. 
A full NLO computation would also require the $O(\alpha_s)$
corrections to the matching conditions. 
Unfortunately, such corrections are not available yet. One might argue
that, being of order $\alpha_s(M_S)$, these contributions should be
small, as suggested by the cases of the SM and of the two Higgs
doublet model;  this statement can only be confirmed by an
explicit computation.  Due to the absence of
$O(\alpha_s)$ corrections to the matching, our ${\cal H}_{\rm
  eff}^{\Delta F=2}$ will be affected by a residual scheme dependence,
which would be cancelled by the missing terms of order
$\alpha_s(M_S)$. \par
A simple  analytic formula for the expression of the Wilson
coefficients as a function of the initial
conditions at the SUSY scale, $C(M_S)$, and of $\alpha_s(M_S)$ can
be given. For 
$M_S > m_t$,  one has 
\begin{equation}
\label{eq:magic1}
C_r(\mu)=\sum_i \sum_s  
             \left(b^{(r,s)}_i + \eta \,c^{(r,s)}_i\right) 
             \eta^{a_i} \,C_s(M_S) 
\end{equation}
where, in the evolution of the coefficients from $M_S$, 
$M_S= (M_{\tilde g} + M_{\tilde q})/2$ has been chosen.
$\eta=\alpha_s(M_S)/\alpha_s(m_t)$, and the magic numbers
$a_i$, $b_i^{r,s}$ and $c_i^{r,s}$ at $\mu= 2$ GeV
in the RI-MOM scheme  can be found in~\cite{silve}.
% are given below:
% \begin{equation}
% \label{eq:magic2}
% \begin{array}{l l}
% a_i=(0.29,-1.1,0.14,-0.69,0.79)& \\ 
% \\ 
% b^{(11)}_i=(0.82,0,0,0,0),& 
% c^{(11)}_i=(-0.016,0,0,0,0),\\ 
% b^{(22)}_i=(0,0.061,0.82,0,0),& 
% c^{(22)}_i=(0,-0.013,0.018,0,0),\\ 
% b^{(23)}_i=(0,-0.092,0,0,0),& 
% c^{(23)}_i=(0,0.013,0.0078,0,0),\\ 
% b^{(32)}_i=(0,-2.9,0.34,0,0),& 
% c^{(32)}_i=(0,0.69,0.012,0,0),\\ 
% b^{(33)}_i=(0,4.4,0,0,0),& 
% c^{(33)}_i=(0,-0.68,0.0055,0,0),\\ 
% b^{(44)}_i=(0,0,0,2.4,-0.003),& 
% c^{(44)}_i=(0,0,0,-0.23,0.0026),\\ 
% b^{(45)}_i=(0,0,0,4.6,-0.38),& 
% c^{(45)}_i=(0,0,0,-1.1,-0.041),\\ 
% b^{(54)}_i=(0,0,0,-0.0024,0.0035),& 
% c^{(54)}_i=(0,0,0,0.00034,-0.0012),\\ 
%% b^{(55)}_i=(0,0,0,-0.0045,0.44),& 
%c^{(55)}_i=(0,0,0,0.0016,0.018) \ ,  \\ 
% \end{array}
%  \end{equation}
% where only  the non-vanishing entries have been written. 
The magic numbers
for the evolution of $\tilde C_{1-3}$ are the same as the ones for the
evolution of $C_{1-3}$. Eq.~(\ref{eq:magic1})
% and (\ref{eq:magic2}) 
can be used in connection with the $B$-parameters,
 given in eq.~(\ref{eq:fres}) below, to compute the
contribution to $\Delta M_K$ and $\varepsilon_K$ at the NLO  for
any model of new physics in which the new contributions with respect
to the SM originate from extra heavy particles. One  has  just to plug
in (\ref{eq:fres}) the $C_i$ of his favourite model. 
%When the $O(\alpha_s)$ corrections to
%the $C_i(M_S)$ are available, one can obtain a full NLO,
%regularisation-independent result; in the cases where this corrections
%have not been computed yet, the results contain a residual systematic
%uncertainty of order $\alpha_s(M_S)$. 
%\par
%The matrix elements of the operators $Q_i$ between $K$ mesons in the
%VIA are given by:
%\begin{eqnarray}
%        \langle K^{0} \vert Q_{1} 
%        \vert \bar{K}^{0} \rangle_{\rm VIA} & = & 
%        \frac{1}{3}M_Kf_{K}^{2}\; ,
%        \nonumber \\
%        \langle K^{0} \vert Q_{2} 
%        \vert \bar{K}^{0} \rangle_{\rm VIA}  & = & -\frac{5}{24}  
%        \left(\frac{M_K}{m_{s}+m_{d}}\right)^{2}M_Kf_{K}^{2}\; ,
%        \nonumber \\
%        \langle K^{0} \vert Q_{3} 
%        \vert \bar{K}^{0} \rangle_{\rm VIA}  & = & \frac{1}{24} 
%        \left(\frac{M_K}{m_{s}+m_{d}}\right)^{2}M_Kf_{K}^{2}\; ,
%        \nonumber \\
%        \langle K^{0} \vert Q_{4} 
%        \vert \bar{K}^{0} \rangle_{\rm VIA} & = & \left[\frac{1}{24} +
%        \frac{1}{4} 
%        \left(\frac{M_K}{m_{s}+m_{d}}\right)^{2}\right]M_Kf_{K}^{2}\; ,
%        \nonumber \\
%        \langle K^{0} \vert Q_{5}
%        \vert \bar{K}^{0} \rangle_{\rm VIA} & = &\left[\frac{1}{8} +
%        \frac{1}{12} 
%        \left(\frac{M_K}{m_{s}+m_{d}}\right)^{2}\right] M_Kf_{K}^{2}\; ,
%        \label{me}
%\end{eqnarray}
% where $M_K$ is the mass of the $K^0$ meson and $m_{s}$ and $m_{d}$ are
% the masses of $s$ and $d$ quarks respectively. 
In the case of the renormalised operators, a convenient definition of the 
$B$-parameters is the following
\bea 
\frac{\langle \bar K^{0} \vert \hat Q_{1} (\mu) 
\vert K^{0} \rangle}{M_Kf_{K}^{2}} &=&  
\frac{B_1}{3}  
\label{eq:bpars} \\
\frac{\langle
\bar K^{0} \vert \hat Q_{2} (\mu) \vert K^{0} \rangle }{M_Kf_{K}^{2}}
&=& -
\left( \frac{ M_K }{ m_{s}(\mu) + m_d(\mu) }\right)^{2}
 \frac{5 B_{2}}{24}\nonumber  \\
\frac{\langle \bar K^{0} \vert \hat Q_{3} (\mu) \vert K^{0} \rangle}
{M_Kf_{K}^{2}}& =&
 \left( \frac{ M_K }{ m_{s}(\mu) + m_d(\mu) }\right)^{2}
\frac{B_3}{24} \nn \\
\frac{\langle \bar K^{0} \vert \hat Q_{4} (\mu) \vert K^{0} \rangle}
{M_Kf_{K}^{2}}& =& 
\left( \frac{ M_K }{ m_{s}(\mu) + m_d(\mu) }\right)^{2}
 \frac{B_4}{4} \nn \\
\frac{\langle \bar K^{0} \vert \hat Q_{5} (\mu) \vert K^{0} \rangle}
{M_Kf_{K}^{2}}& =&
 \left( \frac{ M_K }{ m_{s}(\mu) + m_d(\mu) }\right)^{2}
\frac{B_5}{12}   \nn
\eea 
where the notation $\hat Q_{i}(\mu)$ (or simply $\hat Q_{i}$) denotes
the operators renormalised at the scale $\mu$. Here and in the
following, the same expressions of the $B$-parameters of the operators
$Q_{1-3}$ are valid for the operators $\tilde Q_{1-3}$. 
In eq.~(\ref{eq:bpars}) the operators and the quark masses are
renormalised in the same scheme (RI, $\overline{MS}$, etc.)  at the
scale $\mu$ and the numerical results for the $B$-parameters,
$B_i=B_{i}(\mu)$, presented below refer to the Landau RI scheme.
Moreover, without loss of generality,  terms which are
of higher order in the chiral expansion
% , see eqs.~(\ref{me}), 
and which are usually included in the definition of the $B$-parameters
have been omitted. The 
advantage with the  definition~(\ref{eq:bpars})
 is that the $B$-parameters obeys to very
simple renormalization group equations, with an anomalous dimension matrix 
which is related in a simple way to that of the corresponding 
operators~\cite{dosusy}.
In the  numerical study below, the following values of the $B$-parameters, for 
$\mu=2$ GeV,
have been used
\bea
B_{1} &=& 0.60(6)  \quad
B_{2}= 0.66(4)  \, \nn \\
B_{3} &= &1.05(12) \label{eq:fres}   \\
B_{4} &=& 1.03(6)   \quad 
B_{5} = 0.73(10)  \nn
\eea
The central value  used for $B_1=B_K$ corresponds to $B_K^{\overline{
  MS}}(2\, {\rm GeV})=0.61$, in agreement with the recent estimates 
  of~\cite{Sharpe96}. $B_{2-5}$ have been taken from~\cite{dosusy},
   where all details of the computation can be found
(for another determination of these $B$-parameters, calculated with a
perturbative renormalization, see~\cite{otherb}).
\par 
As an example, I  present the results of a model-independent analysis of 
$K^{0} $--$\bar{K}^{0}$ mixing.
The $K_{L}$--$K_{S}$ mass difference $\Delta M_K$ and the CP-violating
parameter $\varepsilon_K$ are given by
\begin{eqnarray}
        &&\Delta M_K = 2 Re \ \langle K^{0} \vert {\cal
        H}_{\rm eff}^{\Delta S=2} 
        \vert \bar{K}^{0} \rangle \;,
        \label{defdmk}\\
        &&\varepsilon_K = \frac{1}{\sqrt{2} \Delta M_K}
                          Im \  \langle K^{0} \vert {\cal H}_{\rm
                          eff}^{\Delta S=2}   
        \vert \bar{K}^{0} \rangle \;.
        \label{defepsi}
\end{eqnarray}

The SUSY (gluino-mediated) contribution to the low-energy ${\cal
  H}_{\rm eff}^{\Delta S=2}$ contains two real and four complex
unknown parameters: $m_{\tilde q}$, $m_{\tilde g}$,
$\left(\delta^d_{12}\right)_{LL}$, $\left(\delta^d_{12}\right)_{LR}$,
$\left(\delta^d_{12}\right)_{RL}$ and
$\left(\delta^d_{12}\right)_{RR}$.  
The model-independent constraints are obtained by imposing that the
sum of the SUSY contributions proportional to a single
$\delta$-parameter and of the SM contributions to $\Delta M_K$ and
$\varepsilon_K$ does not exceed the experimental value for these
quantities. This is justified  by noting that the constraints on
different $\delta$-parameters in the kaon case exhibit a hierarchical
structure, and therefore interference effects between different
contributions would require a large amount of fine tuning.
The $B$-parameters used in the analysis are those given in 
eq.~(\ref{eq:fres}) 
subtracted by  one standard deviation to their central values, in order to
extract a more conservative bound on SUSY parameters.
The  limits on the $\delta$-parameters for a typical value of the gluino 
and squark masses are  reported in
tables~\ref{tab:reds2_500} and \ref{tab:imds2_500} (more detail can be
found in~\cite{silve}). 
{\scriptsize \begin{table} \centering 
 \begin{tabular}{||c|c|c||}  \hline \hline 
 & NO QCD, VSA  & NLO, Lattice $B_i$  \\ 
 \hline 
 $x$ & \multicolumn{2}{c||}{$\sqrt{|Re  (\delta^{d}_{12})_{LL}^{2}|} $} \\ 
\hline 
 0.3& $1.4\times 10^{-2}$ 

&  $2.2\times 10^{-2}$ 
\\ 
1.0& $3.0\times 10^{-2}$ 
 
& $4.6\times 10^{-2}$ 
\\ 
4.0& $7.0\times 10^{-2}$ 

& $1.1\times 10^{-1}$ 
\\ 
\hline  
 $x$ & \multicolumn{2}{c||}{$\sqrt{|Re (\delta^{d}_{12} )_{LR}^{2}|} 
 \qquad (|(\delta^{d}_{12})_{LR}|\gg 
|(\delta^{d}_{12})_{RL}|)$} \\ 
\hline 
 0.3& $3.1\times 10^{-3}$ 
 
& $2.6\times 10^{-3}$ 
\\ 
1.0& $3.4\times 10^{-3}$

& $2.8\times 10^{-3}$ 
\\ 
4.0& $4.9\times 10^{-3}$ 
 
& $3.9\times 10^{-3}$ 
\\ 
\hline  
 $x$ & \multicolumn{2}{c||}{$\sqrt{|Re (\delta^{d}_{12} )_{LR}^{2}|} 
\qquad ((\delta^{d}_{12})_{LR} = 
 (\delta^{d}_{12} )_{RL})$} \\ 
 \hline 
 0.3& $5.5\times 10^{-3}$

& $1.7\times 10^{-3}$ 
\\ 
1.0& $3.1\times 10^{-3}$

& $2.8\times 10^{-2}$ 
\\ 
4.0& $3.7\times 10^{-3}$

& $3.5\times 10^{-3}$ 
\\ 
\hline  
 $x$ & \multicolumn{2}{c||}{$\sqrt{|Re (\delta^{d}_{12} )_{LL}
(\delta^{d}_{12})_{RR}|} $} \\ 
 \hline 
0.3& $1.8\times 10^{-3}$

& $8.6\times 10^{-4}$ 
\\ 
1.0& $2.0\times 10^{-3}$

& $9.6\times 10^{-4}$ 
\\ 
4.0& $2.8\times 10^{-3}$ 
 
& $1.3\times 10^{-3}$ 
\\ 
\hline  \hline 
 \end{tabular} 
 \caption{Limits on $\mbox{Re}\left(\delta_{ij} 
 \right)_{AB}\left(\delta_{ij}\right)_{CD}$, with $A,B,C,D=(L,R)$, for an
  average squark mass $m_{\tilde{q}}=500$ GeV and for different values
 of $x=m_{\tilde{g}}^2/m_{\tilde{q}}^2$.} 
 \label{tab:reds2_500} 
\end{table} } 
{\scriptsize \begin{table} \centering 
 \begin{tabular}{||c|c|c||}  \hline \hline 
 & NO QCD, VSA & NLO, Lattice $B_i$  \\ 
 \hline 
 $x$ & \multicolumn{2}{c||}{$\sqrt{|Im (\delta^{d}_{12} )_{LL}^{2}|} $} \\ 
\hline 
0.3& $1.8\times 10^{-3}$

& $2.9\times 10^{-3}$ 
\\ 
1.0& $3.9\times 10^{-3}$ 
  
& $6.1\times 10^{-3}$ 
\\ 
4.0& $9.2\times 10^{-3}$ 
  
& $1.4\times 10^{-2}$ 
\\ 
\hline  
 $x$ & \multicolumn{2}{c||}{$\sqrt{|Im (\delta^{d}_{12} )_{LR}^{2}|} 
\qquad (| (\delta^{d}_{12})_{LR}|\gg
 |(\delta^{d}_{12})_{RL}|)$} \\ 
 \hline 
0.3& $4.1\times 10^{-4}$ 
  
& $3.4\times 10^{-4}$ 
\\ 
1.0& $4.6\times 10^{-4}$ 
  
& $3.7\times 10^{-4}$ 
\\ 
4.0& $6.5\times 10^{-4}$ 
  
& $5.2\times 10^{-4}$ 
\\ 
\hline  
 $x$ & \multicolumn{2}{c||}{$\sqrt{|Im (\delta^{d}_{12} )_{LR}^{2}|} 
\qquad ((\delta^{d}_{12})_{LR} = 
 (\delta^{d}_{12} )_{RL})$} \\ 
 \hline 
0.3& $7.2\times 10^{-4}$ 

& $2.2\times 10^{-4}$ 
\\ 
1.0& $4.1\times 10^{-4}$ 
 
& $3.7\times 10^{-3}$ 
\\ 
4.0& $4.9\times 10^{-4}$ 
 
& $4.7\times 10^{-4}$ 
\\ 
\hline  
 $x$ & \multicolumn{2}{c||}{$\sqrt{|Im (\delta^{d}_{12} )_{LL}
(\delta^{d}_{12})_{RR}|}$}  \\ 
 \hline 
0.3& $2.3\times 10^{-4}$ 
 
& $1.1\times 10^{-4}$ 
\\ 
1.0& $2.6\times 10^{-4}$

& $1.3\times 10^{-4}$ 
\\ 
4.0& $3.7\times 10^{-4}$

& $1.8\times 10^{-4}$ 
\\ 
\hline  \hline 
 \end{tabular} 
 \caption{Limits on $\mbox{Im}\left(\delta_{ij}
 \right)_{AB}\left(\delta_{ij}\right)_{CD}$, with $A,B,C,D=(L,R)$, 
 for an average squark mass $m_{\tilde{q}}=500$ GeV and for different values 
 of $x=m_{\tilde{g}}^2/m_{\tilde{q}}^2$.} 
 \label{tab:imds2_500} 
\end{table}}  
\par As a glimpse at tables \ref{tab:reds2_500} and \ref{tab:imds2_500}
readily shows, the inclusion of the NLO QCD corrections to the
Wilson coefficients of ${\cal H}_{\rm
eff}^{\Delta S=2}$ and the use  of the lattice $B$-parameters, instead 
than the
matrix elements evaluated in the vacuum saturation approximation (VSA),
affects the results in different ways, according to the different
operators. The effects are particularly large for left-right operators. 
\section{Non-perturbative renormalization of four-fermion  operators}
\label{sec:npm}
\par  From the theoretical point of view, the  calculation
of hadronic weak decays  can usually be
divided in two parts: the calculation of
the effects of  strong interactions at short
distances, which can be done in perturbation theory using the Wilson
Operator Product Expansion (OPE),
and  the calculation of the hadronic matrix elements of
local operators. The latter contain the effects
of strong forces in the infrared region where
perturbation theory cannot  be applied  and the lattice approach is used.
This is the path followed in  sec.~\ref{sec:susy} 
for the costruction of ${\cal H}_{\rm eff}^{\Delta F=2}$.
An important point in the full approach is the consistency between the 
procedure used to obtain the renormalized operators and the scheme adopted
in the calculation of the Wilson coefficients.
Unlike in many other formulations, as for example in the $1/N$ expansion beyond
the leading order~\cite{bardeen}, and in the chiral quark model~\cite{trieste},
the matching of the renormalized operators to the corresponding Wilson
coefficients can be rigorously done, at least in principle, 
using lattice perturbation theory. 
In practice, the matrix elements  of the renormalized operators constructed 
from the bare lattice ones are subject to two main sources of systematic effects: 
the renormalization constants (matrices) are usually computed
in one-loop lattice perturbation theory and are subject to large 
 ${\cal O}(g^4)$ errors;
 in most of the cases the matrix elements are computed using the
Wilson action and they suffer from  ${\cal O}(a)$ discretization errors.\par 
Numerical studies~\cite{noi}
of the matrix element of the $\Delta S=2$ left-left operator,
$O^{\Delta S=2}= \bar s \gamma_\mu (1-\gamma_5) d \,
 \bar s \gamma_\mu (1-\gamma_5) d$, have shown that 
the chiral behaviour of the renormalized operator is not significantly better
if one adopt a tree-level improved action and operators, whereas it is 
significantly improved by using non-perturbatively determined
renormalization constants (either
by Ward identities~\cite{jlqcd_lat96} or with the non-perturbative method 
(NPM) of~\cite{NPM})~\footnote{ For a different conclusion, using a
boosted improved action and operators, see L. Lellouch at this Conference.}.
\par 
In this section, I review some of the methods which is possible to use
in order to renormalize non-perturbatively the lattice operators,
thus eliminating one of the main sources of uncertainties in the calculation
of the physical amplitudes. 
To this purpose it is useful to start with some definition and classification
of the relevant four fermion operators. In the following
I only discuss the case of $\Delta S=2$ or $\Delta I=3/2$
operators, which cannot mix with operators of lower dimensions.
A full discussion of the renormalization of $\Delta I=1/2$ operators
can be found in~\cite{nlh} and more recently in~\cite{snoi}, see also
\cite{testa}. \par {\bf Operator basis}
In order to illustrate the
renormalization properties of these operators, it is convenient to work in a theory with four
distinct flavours and define the generic operator as 
\bea &\ &O^\pm_{\Gamma_1 \Gamma_2} \equiv \frac{1}{2} \left [
O_{\Gamma_1 \Gamma_2} \pm O^F_{\Gamma_1 \Gamma_2} \right ] =\label{eq:qgamgam}
\\ & \ & \frac{1}{2}
\left [ (\bar\psi_1\Gamma_1\psi_2)(\bar\psi_3\Gamma_2\psi_4) \pm  
(\bar\psi_1\Gamma_1\psi_4)(\bar\psi_3\Gamma_2\psi_2) \right ]
\nn
\eea
where $\Gamma_{1,2}$ denote one of the 16 Dirac matrices,
$ \Gamma
=\{\id,\gamma_{\mu}, \sigma_{\mu\nu}, \gamma_{\mu}\gamma_5,\gamma_5\}
\equiv \{S,V,T,A,P\}$,
and summation of  Lorentz indices  is implied when necessary.
 For example $O_{VV}$ corresponds to $VV\equiv
\sum_\mu \gamma_\mu \otimes \gamma_\mu$, $VA\equiv 
\sum_\mu \gamma_\mu \otimes  \gamma_\mu \gamma_5$, etc. In total we can classify
ten parity-even and ten parity-odd operators.  There are several possible
choices for the operator basis. In the following, I will use 
\bea
Q^\pm_1  &\equiv&  O^\pm_{[VV+AA]}   \quad
Q^\pm_2  \equiv  O^\pm_{[VV-AA]} \nn    \\
Q^\pm_3  &\equiv & O^\pm_{[SS-PP]}  \quad 
Q^\pm_4  \equiv  O^\pm_{[SS+PP]}  \nn \\
Q^\pm_5 & \equiv&  O^\pm_{TT}   \label{eq:qq}
\eea 
for the parity-even operators and 
\bea
\ct{Q}^\pm_1  &\equiv&  - O^\pm_{[VA+AV]} \quad 
\ct{Q}^\pm_2  \equiv  O^\pm_{[VA-AV]} \nn  \\
\ct{Q}^\pm_3 & \equiv & - O^\pm_{[SP-PS]}  \quad 
\ct{Q}^\pm_4  \equiv  O^\pm_{[SP+PS]} \nn \\  
\ct{Q}^\pm_5  &\equiv &  O^\pm_{T \tilde T} 
\label{eq:hh} 
\eea 
for the parity-odd ones. \par 
Parity, chirality and generalized CPS symmetries restrict 
the possible mixing between different  operators~\cite{bernardcps}.
The simplest case is that of the parity-odd operators for which the mixing
matrix,  $\ct{Z}_{ij}$,  is a sparse  block diagonal one. We have
\be \hat \ct{Q}^\pm = \ct{Z}^\pm \ct{Q}^\pm  \ee
where the $\hat \ct{Q}_i$ denote the renormalized operators and 
%{\scriptsize 
{ \begin{equation} \ct{Z}^\pm=
\left(\begin{array}{rrrrr}
\ct{Z}_{11} & 0 & 0 & 0 & 0 \\
0 & \ct{Z}_{22} & \ct{Z}_{23} & 0 & 0 \\
0 & \ct{Z}_{32} & \ct{Z}_{33} & 0 & 0 \\
0 & 0 & 0 & \ct{Z}_{44} & \ct{Z}_{45} \\
0 & 0 & 0 & \ct{Z}_{54} & \ct{Z}_{55}
\end{array}\right)^\pm
\nn
\label{eq:renpv}
\end{equation}}
 The block-structure of the matrix $\ct{Z}_{ij}$ is not changed by
a lattice regularization  which breaks chiral symmetry. 
The situation is rather different for the parity-even operators $Q_i$.
In an ideal, chirally symmetric regularization scheme, $\chi RS$,  the
mixing matrix of the parity-even operators ${Z}_{ij}$
would have the same block-structure
as the mixing matrix of the parity-odd ones.  The explicit chiral symmetry
breaking induces further mixing which can be removed 
by suitable counter-terms which enforce, up to discretization
errors, the relevant Ward identities for the subtracted operators.
A convenient parametrization of the renormalization
matrix is given by the following expressions
\bea
\hat Q^\pm &=& Z^\pm Q^\pm = Z_\chi^\pm \tilde Q^\pm \\
\tilde Q^\pm &=& [I+\Delta^\pm]Q^\pm
\eea
where 
%{\scriptsize 
{\be Z^\pm_\chi=
\left(\begin{array}{rrrrr}
Z_{11} & 0 & 0 & 0 & 0 \\
0 & Z_{22} & Z_{23} & 0 & 0 \\
0 & Z_{32} & Z_{33} & 0 & 0 \\
0 & 0 & 0 & Z_{44} & Z_{45} \\
0 & 0 & 0 & Z_{54} & Z_{55}
\end{array}\right)^\pm
\label{eq:renor_subt} \nn
\end{equation}} 
and 
%{\scriptsize 
{\begin{equation} \Delta^\pm = 
\left(\begin{array}{rrrrr}
0 & \Delta_{12} & \Delta_{13} & \Delta_{14} & \Delta_{15} \\
\Delta_{21} & 0 & 0 & \Delta_{24} & \Delta_{25} \\
\Delta_{31} & 0 & 0 & \Delta_{34} & \Delta_{35} \\
\Delta_{41} & \Delta_{42} & \Delta_{43} & 0 & 0 \\
\Delta_{51} & \Delta_{52} & \Delta_{53} & 0 & 0
\end{array}\right)^\pm
\label{eq:renpc_sub}\nn
\end{equation} }
%The renormalization of the parity
%conserving sector is given by $Z^\pm = Z_\chi^\pm [I+\Delta^\pm]$.
Note that in the hypothetical $\chi RS$, $\Delta^\pm=0$
and $\ct{Z}_{ij}=Z_{ij}$.
\par  {\bf Ward identities and the NPM}
I now discuss, with one simple example, the determination
of the mixing matrix $\Delta^\pm$  using the Ward identity method (WIM).
The WIM, first discussed  in \cite{boc} (see also \cite{lucgui}), was  
applied in \cite{ZETA_A} 
to determine the renormalization constant of the axial current using 
quark Green functions and then implemented 
to the non-perturbative determination of the mixing
coefficients of the $\Delta S=2$ operator $Q_1$ in~\cite{jlqcd_lat96}.
I will also show that the WIM is equivalent to the NPM on quark and gluon states,
first proposed in~\cite{NPM} and then implemented for the renormalization
of the four-fermion operators in~\cite{noi}. A more detailed discussion
can be found in~\cite{neverendingstory}.
Before discussing the Ward identities, I have to introduce some definition. 
The 
Green functions considered in the following are given by the  expectation values of 
the following multilocal  operators: 
%{\scriptsize 
{\bea
\nonumber
G_k (x_0;x_{1,2,3,4})&=&
\psi_1\bar \psi_2
Q_k(x_0) \psi_3\bar \psi_4\\
\ct{G}_k (x_0;x_{1,2,3,4}) &=&
\psi_1 \bar \psi_2\ct{Q}_k(x_0)
\psi_3\bar \psi_4\nonumber\\
\nonumber
\ct{G}_k^{(1)} (x_0;x_{1,2,3,4})
&=& [\gamma_5 \psi_1] \bar \psi_2 \ct{Q}_k(x_0)
\psi_3 \bar \psi_4 \\
\ct{G}_k^{(2)} (x_0;x_{1,2,3,4})
&=& \psi_1[\bar \psi_2 \gamma_5] \ct{Q}_k(x_0)
\psi_3 \bar \psi_4 \nn
\label{eq:defops}
\eea } 
and similarly for $\ct{G}_k^{(3)}$ and $\ct{G}_k^{(4)}$; $k=1,\dots,5$.
In the above equation $\psi_1=\psi_1(x_1)$, $\psi_2=\psi_2(x_2)$, etc.
For simplicity, the colour and spin indices have not been
shown explicitly and the
$\pm$ superscripts have  been dropped  from operators and correlation functions.
Having thus defined $G_k$,
$\ct{G}_k$ $\ct{G}_k^{(1)}$ etc. in coordinate space, we will also be using
the corresponding Green functions in momentum space $G_k(p)$,
$\ct{G}_k(p)$, $\ct{G}_k^{(1)}(p)$ etc. (all external legs having
the same Euclidean momentum). The amputated Green functions of
$\langle G_k(p) \rangle$, $\langle \ct{G}_k(p) \rangle$ and
$\langle \ct{G}_k^{(1)}(p)\rangle$ will be denoted by $\Lambda_k(p)$,
$\ctLambda_k(p)$ and $\ctLambda_k^{(1)}(p)$, respectively.
\par On quark states, the relevant Ward identity for the operator $Q_1$ can be written as
\bea
&& \langle G_1 \rangle
-\frac{1}{4}\left( \langle \ct{G}_1^{(1)} \rangle
- \langle \ct{G}_1^{(2)} \rangle
+ \langle \ct{G}_1^{(3)} \rangle
- \langle \ct{G}_1^{(4)} \rangle\right) = \nonumber \\
&& - \int d^4 x \langle \ct{G}_1
\left[ \nabla^\mu A_\mu(x) - 2 m_0 P(x) - X_5(x) \right] \rangle \nn \\
\label{eq:wi1}
\eea 
where the flavour indices of $A_\mu(x)$ and $P(x)$ have been omitted. 
Out of 
the chiral limit, the $2 m_0 P(x)$ term is present, but the surface term
from the current divergence vanishes upon integration. I will be considering
this case, in order to mimic what is practicaly done in numerical
simulations
(i.e. first we compute quantities at small non-zero quark mass and then extrapolate
to the chiral limit). 
The operator $X_5$ arises from the variation of the chiral symmetry breaking
Wilson term in the action. As shown in~\cite{boc} (see also
\cite{crluvl} for a detailed discussion) it mixes, under
renormalization with $\nabla^\mu A_\mu$ and $P$. This mixing, determined
by the requirement that on-shell matrix elements of the subtracted $X_5$
vanish in the continuum, generates a finite renormalization of the axial current
and a power subtraction of the quark mass. Thus, following~\cite{boc}, in the above WI we will trade-off $X_5$ for the
renormalized expression $[\nabla^\mu \hat A_\mu - 2 \hat m \hat P -
\overline X_5]$, where $\overline X_5$ is the subtracted $X_5$.
What is of interest to us is that, besides the above renormalizations, the 
$\overline X_5$ insertion in the above corelation function also generates
new contact terms 
\bea
&&\langle \overline X_5(x) \ct{G}_1(x_0;x_1,x_2,x_3,x_4) \rangle =\nn \\
&&(1 - \frac{Z_{11}}{\ct{Z}_{11}}) \langle G_1 \rangle \delta(x-x_0)\nn\\&&
- \frac{Z_{11}}{\ct{Z}_{11}} \sum_{k=2}^5 \Delta_{1k} \langle G_k \rangle
\delta(x-x_0)
\label{eq:xct}
\eea 
where the notation for the various coefficients has been chosen with some
foresight.  There are also contact terms
arising from the proximity of $\overline X_5(x)$ to the quark fields
of the correlation $\ct{G}_1$ at points $x_1,\cdots,x_4$. These 
terms have the form $\langle \ct{G}_1^{(k)} \rangle \delta (x-x_k)$
(with $k=1,\cdots,4$). However, as shown in~\cite{testa},
they vanish in the continuum limit.
We now combine eqs.~(\ref{eq:wi1}) and (\ref{eq:xct}), Fourier-transform
the WI (with all external momenta set equal to $p$) and amputate the
resulting correlation functions 
\bea && \tilde \Lambda_1 (p) =
\Lambda_1 (p) + \sum_{k=2}^5 \Delta_{1k} \Lambda_{1k} (p) = \nn \\ &&
\frac{\ct{Z}_{11}}{4 Z_{11}} [\ctLambda _1^{(1)} - \ctLambda _1^{(2)}
+ \ctLambda _1^{(3)} - \ctLambda _1^{(4)} ] \nonumber \\
&& + \frac{\ct{Z}_{11}}{Z_{11}} \int d^4 x \langle \ct{G}_1 (p)
\hat m \hat P(x) \rangle \prod_{j=1}^4 \langle S^{-1}(p) \rangle 
\label{eq:wi1mom}
\eea 
We require  the above WI, up to quark field renormalization, to
be the one valid in the continuum limit for subtracted correlation functions
(operators).  This implies
that the l.h.s., multiplied by $Z_{11}$, can be identified with the renormalized parity-conserving
correlation function (operator). We see immediately that the WI fixes the
operator subtractions (i.e. the mixing coefficients $\Delta_{1k}$).
This can be done by projecting the above WI with suitable projectors~\cite{neverendingstory}
and by solving the resulting system of linear non-homogenous equations for the
 four $\Delta$'s~\footnote{ The projectors 
obey the following orthogonality conditions:
$
\Tr\Pj^\pm_i \Lambda^{\pm (0)}_k = \delta_{ik}$ and  
$\Tr \Pjpv ^\pm_i {\ctLambda}^{\pm (0)}_k =
\delta_{ik}$ with  $i,k = 1,\dots,5$ 
where $\Lambda^{(0) \pm}_k$ and $\ctLambda^{(0) \pm}_k$ are the tree-level
amputated Green functions of $Q^\pm_k$ and ${\ct{Q}}^\pm_k$
respectively.}.
The WI also fixes the ratio of the multiplicative
renormalization constants $Z_{11}/\ct{Z}_{11}$, but not each of them
separately.  This is the analog of what happens in the case of the scalar and
pseudoscalar densities: the Ward identities can only fix the ratio of
the renormalization constants of these operators, $Z_P/Z_S$, but not $Z_P$ or
$Z_S$. Their values, for fixed $Z_P/Z_S$,
 is arbitrary and depend on the renormalization scheme.
\par The WI (\ref{eq:wi1mom}) is only a function of  bare 
quantities. This
demonstrates that the matrix $\Delta$ and the ratio $Z_{11}/\ct{Z}_{11}$
are  finite functions of  $g_0^2$, which can be
completely determined from the WIs, and do not depend on the 
renormalization conditions. The overall renormalization constant
$Z_{11}$ ($\ct{Z}_{11}$), instead, is logarithmically divergent (as in the continuum) 
and is  fixed by imposing a suitable (scheme-dependent) renormalization condition
on the subtracted operator $\tilde Q_1$.  
\par I will  demonstrate now  that the WIM and the NPM
 are equivalent for the determination of the
lattice mixing coefficients. The validity of this statement has 
already been discussed in general in~\cite{NPM} and in the specific 
case of   the four-fermion operators in~\cite{testa}.
Here I discuss the specific example of $Q_1$. It
is to be understood, throughout the rest of this section, that the chiral
limit is to be taken in the end.
\par From (\ref{eq:wi1mom}), in the large momentum limit, one may derive 
the  renormalization conditions used in the NPM to fix the matrix $\Delta$.
For large $p$, the last term on the r.h.s. of eq.~(\ref{eq:wi1mom})
is power supressed. This happens because the explicit $m$ factor implies that the
integrand has one less dimension that the other terms so that, at large
momenta, it vanishes faster by one power of $p$.
Moreover, in this limit
the inverse quark propagator has the general form $S^{-1}(p) = i \Sigma_1
\gamma_\mu p_\mu$ (with $\Sigma_1$ a scalar form factor). This means that in
this limit it anticommutes with $\gamma_5$. By projecting  both sides of 
eq.~(\ref{eq:wi1mom})
 on the single operators
$Q_1$, $\Pj_k$ (with $k=1,\dots,5$)~\cite{neverendingstory},  one  obtains  
\be
 \left[ \Tr \Pj_j \Lambda_1 + \sum_{k=1}^5 \Delta_{1k} \Tr \Pj_j
\Lambda_k \right] = \frac{\ct{Z}_{11}}{Z_{11}} \Tr \Pjpv _j \ctLambda _1 
\ee 
which is precisely the system of equations used in~\cite{noi} and \cite{neverendingstory} to fix the coefficients of the
 mixing matrix $\Delta$.
\par The generalization to  the other operators is straighforward, although
more complicated. For example, considering the 
operators $Q_{2,3}$ and $\ct{Q}_{2,3}$, we have  
\bea
[\Lambda (p) + \Lambda_\Delta (p) z_\Delta^T ] z_\chi^T 
= -\frac{1}{2} [\ctLambda^{(1)} - \ctLambda^{(3)}] \zeta_\chi^T
\nonumber \\
+ \int d^4 x \langle \ct{G} (p) \zeta_\chi^T
2 \hat m \hat P(x) \rangle \prod_1^4 \langle S^{-1}(p) \rangle
\label{eq:wi2mom}
\eea 
where I have used a compact notation
for the row-vectors $\ct{G} \equiv (\ct{G}_2,\ct{G}_3)$, $\Lambda\equiv 
(\Lambda_2,\Lambda_3)$, 
$\ctLambda\equiv (\ctLambda_2,
\ctLambda_3)$,  $\Lambda_\Delta\equiv 
(\Lambda_1,\Lambda_4,\Lambda_5)$ 
and for the  mixing  matrices
\bea
\begin{array}{c} 
\zeta_\chi
\end{array}
=
\left(\begin{array}{rr}
\ct{Z}_{22} & \ct{Z}_{23} \\
\ct{Z}_{32} & \ct{Z}_{33}
\end{array}\right)
\nonumber \\
\begin{array}{c} 
z_\chi
\end{array}
=
\left(\begin{array}{rr}
Z_{22} & Z_{23} \\
Z_{32} & Z_{33}
\end{array}\right)
\nonumber \\
\begin{array}{c} 
z_\Delta
\end{array}
=
\left(\begin{array}{rrr}
\Delta_{21} & \Delta_{24} & \Delta_{25} \\
\Delta_{31} & \Delta_{34} & \Delta_{35}
\end{array}\right)
\eea 
Equation~(\ref{eq:wi2mom}) correspond to  a system of linear equations in 
the matrix elements of $z_\Delta$ and in $z_\chi \ct\zeta_\chi^{-1}$ 
which can be used to fix the subtracted operators $\tilde Q_{2,3}$. 
Finally,  overall renormalization conditions have to be imposed to the 
$\tilde Q_{2,3}$ to obtain the $\chi RS$ mixing matrix $z_\chi$.
\par Finally, I want to mention another method,
 implicitly suggested in \cite{boc,snoi}, which has not been implemented
yet in numerical simulations.  This method is based on the Ward identities
which can be written for gauge-invariant correlation functions.
Let us introduce the  ``meson fields"  $P^K(t_K)=\int d^3x \bar s(\vec x,t_k)
\gamma_5 d(\vec x, t_k)$, $P^{\pi+}(t_\pi)=\int d^3x \bar u(\vec x,t_\pi)
\gamma_5 d(\vec x, t_\pi)$, etc.
By defining the following correlation functions 
\bea && G_2(t_{\pi_1},t_{\pi_2})= \langle 0\vert P^{\pi_1}(t_{\pi_1})
\tilde Q(x_0) P^{\pi_2}(t_{\pi_2})\vert 0 \rangle \nn \\ &&
\ct{G}_3(t_{\pi_1},t_{\pi_2})= \int dt_K  \langle 0\vert P^{K}(t_{K})
\tilde \ct{Q}(x_0)\nn \\ &&
 P^{\pi_1}(t_{\pi_1}) P^{\pi_2}(t_{\pi_2})\vert 0 \rangle 
\eea 
the Ward identity can schematically be  written as 
\be \lim_{m\to 0} 2 m \ct{Z} \ct{G}_3(t_{\pi_1},t_{\pi_2})
= const. \times Z G_2(t_{\pi_1},t_{\pi_2}) \label{eq:giwi} \ee 
where $const.$ is a suitable constant which depends on the flavor quantum
numbers of the operators $Q$ and $\ct{Q}$ and of the external 
``meson fields". For different values of $t_{\pi_1}$ and $t_{\pi_2}$,
the above Ward identity results in  a system of linear 
non-homogenous equations from which it 
is possible to extract $\Delta^\pm$  and $Z_\chi \ct{Z}^{-1}_\chi$.
A study in this direction is underway~\cite{bouc}. 
\section{OPE without OPE}
\label{sec:opewope}
In this section, I will discuss some new methods which have been developed in order
to overcome the difficulties related to the renormalization of lattice composite
operators relevant in many important physical applications.
Among the others, let me mention the structure functions in deep inelastic
scattering, kaon and $B$-meson non-leptonic exclusive  decays, electromagnetic
form factors at large momentum transfer,  exclusive and inclusive
semileptonic $B$-decays etc. The general idea underlying these new methods
is essentially the same and I will denote it as Operator Product Expansion
without Operator Product Expansion (OPE without OPE or OPEwOPE). The reason for choosing this
name will be clearer in the following. I now discuss the general features
of the OPEwOPE, starting from the description of the standard approach 
followed so far in lattice calculations.
\par Let us consider the short-distance 
OPE  for the $T$-product of two generic currents between the 
physical external states $\vert A \rangle$ and $\vert B \rangle$ 
\bea \langle B \vert T ( J(x) J(0)  \vert A \rangle \equiv
\sum_i C_i(x,\mu) \langle B \vert \hat O_i(\mu)  \vert A \rangle 
%\nn \\
 \label{eq:ope}\eea 
where, for simplicity, Lorentz indices have been omitted. In the above equation,
$\mu$ is the renormalization scale of the
local, composite operators $\hat O_i(\mu)$ and
the $C_i(x,\mu)$ are the Wilson coefficients,  which can be computed in perturbation
theory, provided  $\mu \gg \Lambda_{QCD}$. The Wilson coefficients are choosen
in such a way that the $T$-product is $\mu$ independent. The standard approach
  consists in computing the
Wilson coefficients in perturbation theory and   extracting the matrix elements of the
local operators from suitable hadronic correlation functions. 
Many examples of matrix elements of local operators
that have been studied  in the past can be given:
i) Moments of the structure functions, e.g.
$ \langle p \vert
 \bar \psi_i \gamma_{\mu_1} 
i \dlr_{\mu_2} \dots  i \dlr_{\mu_n} \psi\vert p \rangle $; 
ii) Hadronic weak decays ($K \to \pi \pi$ or $B \to \pi \pi$), e.g. 
$\langle \pi \pi \vert \left( \bar s \gamma_\mu^L d \right)\left(
\bar u \gamma^\mu_L e \right) \vert K \rangle $;
iii) $K^0$--$\bar K^0$ and $B^0$--$\bar B^0$ mixing, e.g.
$\langle \bar K^0 \vert \left( \bar s \gamma_\mu^L d \right)
\left(\bar s \gamma^\mu_L d \right) \vert K^0 \rangle$;
iv) operators of the HQET entering inclusive and exclusive decay
rates, e.g.
 $\langle B \vert \bar h_v D_0 h_v \vert B \rangle$ and
$\langle B \vert \bar h_v ( i \dr)^2 h_v \vert B \rangle$;
iv) light-cone wave functions for  electromagnetic
form factors and exclusive $B \to \pi$ and $B \to \rho$ semileptonic
decays, e.g.  $\langle \pi \vert \bar \psi_i \gamma_{\mu_1} \gamma_5 
i \dlr_{\mu_2} \dots i \dlr_{\mu_n} \psi \vert 0 \rangle$.
\par
The renormalized
operators $\hat O_i(\mu)$ are obtained from the bare lattice operators
either by using (boosted-tadpole improved) perturbation theory
or non-perturbatively using the WIM or the NPM.  The difficulties encountered in 
this approach are due to the fact that the ultraviolet behaviour
of the ``effective theory" (OPE, HQET, NRQCD, etc.) is much worse and
divergences which were not present in the ``full" theory appear in the
different expansions.
For example, if the currents are the SM  quark currents responsible
for weak decays, the  l.h.s. of eq.~(\ref{eq:ope}) is finite, whereas the operators
appearing on the r.h.s. are logarithmicaly or power divergent in the
inverse lattice spacing $a^{-1}$.  We briefly discuss examples taken from the 
cases considered above.
\par i) Given the bare energy-momentum tensor
density  $T^{\mu\nu}= \bar \psi (\gamma_\mu \dlr_\nu +
\gamma_\nu \dlr_\mu ) \psi$, the corresponding renormalized operator is given by
\bea &&  \hat T^{\mu\nu}(\mu) = Z(\mu) \left[ 
%\bar \psi \left(\gamma_\mu \dlr_\nu +
%\gamma_\nu \dlr_\mu \right) \psi \right.
%\nn \\ && \left. 
T^{\mu\nu}
 + \frac{C(g_0^2)}{a} \delta_{\mu\nu}
\bar \psi \psi \right]  \eea
In this case it is easy to get rid of the power divergences by taking
$\mu \neq \nu$.
\par ii)  In presence of a charm-up
 GIM mechanism,  the renormalized $O^\pm$ operators have the
form~\cite{boc,nlh} 
\bea & & \hat O^\pm(\mu) = Z^\pm(\mu) \left[ O^\pm
%\left(\bar s \gamma_\mu^L d  \right)
%\left(\bar u \gamma^\mu_L u \right) \pm \left( \bar s \gamma_\mu^L u  \right)
%\left( \bar u \gamma^\mu_L d \right)\nn \right. \\ && \left.
+ \sum_i C_i^\pm(g_0^2) O_i
\right. \nn  \\ & &  \left. + (m_c-m_u) C_\sigma^\pm(g_0^2) 
\bar s \sigma^{\mu\nu} G^{a}_{\mu\nu} t^a d  \right.  \nn \\ & &
\left. 
+ (m_c-m_u) (m_s-m_d) \frac{C^\pm_P(g_0^2)}
{a} \bar s \gamma_5 d  \right. \nn \\&& \left. + (m_c-m_u)  \frac{C^\pm_S(g_0^2)}
{a^2} \bar s d \right]   \eea 
where the  $O_i$ are operators of dimension
six, the mixing of which is induced by the explicit
chiral symmetry breaking  of the lattice action and have been 
discussed in sec.~\ref{sec:npm}.
In this case, it has been impossible so far to get the
matrix elements of the renormalized operators with  resonable accuracy.
\par iii)  By defining $S_0= \bar h_v D_0 h_v  $ and
$K= \bar h_v ( i \dr)^2 h_v $,
the renormalized operators have the form 
\bea  &&  \hat S_0(\mu) = Z_h(\mu) \Bigl( \bar h_v D_0 h_v +\frac{C_h(g_0^2)}{a}
\bar h_v h_v \Bigr)  \nn \\ && 
\hat K(\mu) = Z_{D^2}(\mu) \Bigl( \bar h_v ( i \dl)^2 h_v  
\\ && +
\frac{C_1(g_0^2)}{a} \bar h_v D_0 h_v +\frac{C_2(g_0^2)}{a^2}
\bar h_v h_v \Bigr)  \nn \eea 
The subtraction procedure necessary to obtain a finite $\hat S_0$, which 
is related to the definition of the renormalized  quark mass, 
is straightforward~\cite{cgms}. An accurate determination
of the quark mass, however,  
 requires the control of high orders in perturbation
theory~\cite{msren}; the definition of a finite $\hat K$, necessary
to obtain the heavy quark kinetic energy is much harder and a
small error on the  final answer  difficult to obtain~\cite{cgms}. 
\par The use of perturbation theory in the calculation of the mixing coefficients necessary
to subract power divergencies gives unreliable results. A  detailed discussion, with several examples can be 
found in~\cite{msren}. For the calculation of logarithmicaly divergent or finite mixing 
coefficients instead,
perturbation theory can be used, at least in principle. In practice, however, due to large 
higher-order corrections present in the expansion of the lattice perturbative
series, the results are not very accurate and it is for this reason that the non-perturbative methods discussed 
in sec.~\ref{sec:npm} have been developed.
\par All the problems connected to the construction of finite operators, out of the bare (power) divergent 
lattice ones, can be overcome, at least in principle, by using the OPEwOPE. I  start by recalling a result
presented  at the Latt97~\cite{snoi}, and then  discuss
new proposals, and feasibility studies, which appeared this year.
\par {\bf Weak Hamiltonian}
 The first example of OPEwOPE is related to non-leptonic weak decays.
The standard construction of the weak Hamiltonian
begins with the  expression  
\begin{equation}
\frac{g^2_{W}}{2}\int d^4x\,D_{\rho\nu}^W(x;M_{_W}) 
T\left[J_{\rho L}(x) J^\dagger_{\nu L}(0)\right] 
\label{eq:HEFF}
\end{equation} 
where 
$D_{\rho\nu}^{W}(x;M_{_W})$ 
is the $W$-boson propagator and $J_{\rho L}$ is the (left-handed)
hadronic weak current.  One then performs the OPE of the product of the 
two currents in eq.~(\ref{eq:HEFF}) 
\bea &&
\langle B |{\cal H}_{\rm eff}^{W}| A\rangle = \nn \\ &&
\frac{G_{F}}{\sqrt{2}} \sum_i C_i(\mu,M_{_W}) M_{_W}^{6-d_i} 
\langle B |\hat{O}^{(i)}(\mu)| A \rangle
\label{eq:HEFFOPE}
\eea 
where $d_i$ is the dimension of the operator $\hat{O}^{(i)}(\mu)$,
and the functions $C_i(\mu,M_{_W})$ result from the integration
of the Wilson expansion coefficients, $c_i(x;\mu)$ (defined in 
eq.~(\ref{eq:ME}) below), with the
$W$-propagator. Schematically, suppressing Lorentz indices, one has 
\begin{equation}
\frac{C_i(\mu,M_{_W})}{ M_{_W}^{d_i-6}} 
= \int d^4x\,D^{W} (x;M_{_W}) c_i(x;\mu)
\label{eq:WILCOEF}
\end{equation} 

The sum in the expansion~(\ref{eq:HEFFOPE}) is over operators of
increasing dimension. In the following only operators
with dimensions $d_i \le 6$ are considered, since the contribution from operators
with $d_i>6$ is suppressed by powers of $1/M_{_W}$.

All the intricacies of operator mixing in the definition of the finite
and renormalized operators, $\hat{O}^{(i)}(\mu)$, come about because the
integral in~(\ref{eq:HEFF}) is extended down
to the region of extremely small $x$. The complicated mixing for the
$\hat{O}^{(i)}(\mu)$'s in terms of bare operators arises from contact
terms when the separation of the two currents goes to zero (i.e. when
$|x|$ is of the order of $a$).  This
observation suggests that a simple way to avoid these complications is
to  define the renormalized operators by enforcing the
OPE  for distances $|x|$ much larger than the lattice
spacing $a$. This is done by computing directly the
$T$-product of the two currents rather than the matrix elements
of the local operators on the r.h.s. of eq.~(\ref{eq:HEFFOPE}). For
this reason I call this method OPEwOPE.
We imagine proceeding in the following way:
1)
Take the $T$-product of two properly normalized weak currents,
$J_{\rho L}(x) J^\dagger_{\rho L}(0)$. If required these currents can
 be improved.
2)
Measure the hadronic matrix element
$\< B \vert T[J_{\rho L}(x) J^\dagger_{\rho L}(0)]\vert A \>$
in a Monte Carlo simulation,
as a function of $x$ for $|x|\rightarrow 0$ in the region 
$
a\ll |x|  \ll  \Lambda_{QCD}^{-1} 
$.
3)
Extract the numbers $\< B \vert \hat{O}^{(i)}(\mu)\vert A \>$ by
fitting in the region defined above the $x$-behaviour of 
$\< B \vert T[J_{\rho L}(x) J^\dagger_{\rho L}(0)]\vert A \>$ to the formula 
\bea && 
\< B \vert T\left[J_{\rho L}(x) J^\dagger_{\rho L}(0)\right]\vert A \> =
\nn \\ && \sum_i
c_i(x;\mu) \< B \vert {O}^{(i)}(\mu)\vert A \> 
\,,\label{eq:ME}
\eea 
where the Wilson coefficients $c_i(x;\mu)$ are determined by
continuum perturbation theory using any standard renormalization scheme.
 Since 
we only consider operators of dimension 6 or lower, the $T$-product differs
from the right-hand side of eq.~(\ref{eq:ME}) by terms of $O(|x|^2\lqcd^2)$,
which is an estimate of the size of the systematic errors in this procedure.
4)
Insert the numbers $\< B \vert \hat{O}^{(i)}(\mu)\vert A\>$ 
determined in this way into the expression for the 
matrix elements of  ${\cal H}_{\rm eff}^{W}$ in  
eq.~(\ref{eq:HEFFOPE}).

For the implementation of this procedure, what is required is the
existence of a window, in which the distance
between the two currents is small enough, so that perturbation theory
can be used to determine the expected form of the OPE, but large
enough that lattice artifacts are small.  
The existence of a window is a necessary conditions in all applications
of the OPEwOPE. 
\par 
The method determines directly the ``physical'' 
matrix elements of the operators appearing in the OPE of the 
two currents, i.e. the matrix elements of the finite, renormalized operators
$\hat{O}^{(i)}(\mu)$, without any reference to the magnitude of the $W$-mass. 
Thus we do not need to probe distances of $O(1/M_{_W})$ with lattice
calculations.
The $\mu$-dependence of the matrix elements of the operators
$\hat{O}^{(i)}(\mu)$ is given trivially by that of the (perturbative)
Wilson coefficients, $c_i(x;\mu)$.  It compensates the related
$\mu$-dependence of the functions $C_i(\mu,M_{_W})$ in such a way that
the l.h.s of eq.~(\ref{eq:HEFFOPE}) is independent of the choice of the
subtraction point. A similar comment holds for the dependence on
renormalization scheme. \par
{\bf Non-linear $\sigma$-model}
 A feasibility study  of this method,  in a simplified case,
has be presented by S.~Caracciolo in a parallel session at this
 Conference~\cite{cmp}. 
They have considered the two-dimensional non-linear $\sigma$-model, 
for which it is
possible to define the following conserved currents 
\be j_\mu^{a,b}(x) = \frac{1}{g} \left( \sigma^a(x) \partial_\mu \sigma^b(x)
-  \sigma^b(x) \partial_\mu \sigma^b(x) \right) \ee 
where $\vec \sigma \equiv (\sigma^1, \dots, \sigma^N)$ with the condition
$\vec \sigma(x) \cdot \vec \sigma(x)=1$.
The relevant expression is 
\be \langle p \vert \sum_{a,b} j_\mu^{a,b}(x) j_\nu^{a,b}(0) \vert p \rangle
= C_T(x; \mu ) \langle p \vert  \hat T_{\mu\nu} \vert p \rangle \, 
\label{eq:cara}\ee 
where $\vert p \rangle$ is some external state of momentum $p$ and
$\hat T_{\mu\nu}$ the renormalized  energy-momentum tensor
($T_{\mu\nu}= \partial_\mu \vec \sigma \cdot \partial_\nu \vec \sigma$). This operator
has zero anomalous dimension, but it is subject to a finite renormalization
(analogous to $Z_{V,A}$ in the case of Wilson fermions)
on the lattice. \par 
In the OPEwOPE approach, one computes non-perturbatively the matrix elements
of the $T$-product of the original currents, corresponding to the l.h.s
of (\ref{eq:cara}) and the Wilson coefficients (in this case there
is only one operator  corresponding to 
$C_T(x;\mu)$) in perturbation theory. From the knowledge of these quantities,
one can then extract the desired matrix element of the local operators,
i.e. $\langle p \vert  \hat T_{\mu\nu} \vert p \rangle$.
In the example considered here, we may also compute  directly
the matrix element of $\hat T_{\mu\nu}$ and check to what extent the
procedure works.  In~\cite{cmp} they found a large window of distances
for which the OPE is verified, when continuum perturbation theory
is used in the calculation of the Wilson coefficient,
giving the correct matrix element
independently of the external states.
\par {\bf The quark propagator} The simplest $T$-product that can be considered
is the quark propagator in a fixed gauge, for example the Landau gauge.
Let us consider for example the following quantity 
\bea \Sigma_2(x) = \langle 0 \vert \psi(x)^\alpha_A \bar 
\psi(0)^\alpha_A \vert 0 \rangle \, ,\eea 
where $\alpha$ and $A$ are spin and color indices respectively. 
At short distances, i.e. as $\vert x \vert \to 0$, we have 
\bea && \Sigma_2(x) \to C_\psi(x ; \mu) \langle \bar \psi \psi (\mu) \rangle
+ C_m(x; \mu) \frac{\hat m(\mu)}{\vert x\vert^2}\nn \\ && +
{\cal O}\left(\frac{a}{\vert x \vert^4}\right)  \label{eq:s2x} \eea 
where $C_\psi(x; \mu)$ and $C_m(x; \mu) $ are the Wilson coefficients,
$ \langle \bar \psi \psi (\mu) \rangle$ the quark condensate and
${\cal O}(a/\vert x \vert^4)$ represent (gauge non-invariant)
lattice artifacts~\cite{unpu}.  The Wilson coefficients, which are
gauge-dependent, can be computed in any continuum renormalization scheme,
for example in the $\overline{MS}$ scheme~\cite{PdR}. 
In the chiral limit, by fitting $\Sigma_2(x)$ to
(\ref{eq:s2x}),   and using $C_\psi(x; \mu)$ and  $C_m(x; \mu) $
from perturbation theory, one can extract the value of the condensate
 and the renormalized
quark mass in the same scheme as the Wilson coefficients. It would be
very interesting to compare the values obtained in this way, with those
obtained with other techniques. Note that if we use the Fourier transform
of $\Sigma_2$, there are further contributions from contact terms, which 
can make the extraction of the condensate and of the quark mass more
complicated.
\par {\bf The shape function}  The knowledge of the shape function
$f(k_+)$ is a fundamental
ingredient for the
extraction of $\vert V_{ub} \vert$ from the end-point of the lepton spectrum.
The same function also enters the calculation of the photon spectrum in
radiative $B$ decays. 
To give an explicit example, the differential distribution 
in semileptonic decays is given by 
\bea \label{eq:dslr} &&
\frac{d \Gamma}{dE_{\ell}} \equiv \int^{M_B}_{0} dk_{+} \ f(k_{+}) \
\frac{d \Gamma_{PM}}{dE_{\ell}}(m_b^*,E_\ell)\nn \\ && = \vert V_{u b} 
\vert^{2} \frac{G_{F}^{2} }{12 \pi^{3} }\ E_{\ell}^{2} 
 \int^{M_B}_{0} dk_{+} \ f(k_{+}) \nn \\ &&
\Theta( m^{*}_{b} - 2 E_{\ell} ) \left[ 3 m^{* \, 2}_{b} - 4 m^{*}_{b} E_{\ell} 
\right] \, , \eea 
 where $m^{*}_{b}= M_B- k_{+}$.
\par The physical rate can be derived from the imaginary
part of the forward matrix element
of the $T$-product of two weak currents 
\bea && W^{\mu \nu} = \frac{1}{\pi} \mbox{Im}  i \int \ d^4x  \ 
 e^{-i q \cdot x} \times \nn \\ &&  \ 
  \langle \bar B(v) \vert T \Bigl[ J^{\mu \ \dagger} (x)\  J^\nu(0)\Bigr]
\vert \bar B(v)  \rangle  \eea 
In the standard approach, one applies the OPE to the $T$-product above,
and $W^{\mu\nu}$ is written in terms of the following matrix elements
of local operators 
\be {\cal M}_n  \propto\langle \bar B(v) \vert \bar b_v \gamma^\nu (i D^{\mu_1}) \dots
(i D^{\mu_n}) b_v \vert \bar B(v) \rangle 
\ee 
which correspond to the moments of the shape function
\be  \label{eq:mom} {\cal M}_{n} = \int_{0}^{M_B} 
dk_{+} k^{n}_{+} f(k_+) \eeq
With the OPEwOPE, it is possible to obtain the full
shape function. The final result of~\cite{aglie1}
is very simple. One study the ratio of two 
correlation functions  
\bea \label{eq:rw}&& W^{\mu\nu}(t, \vec Q)= \\
&& \lim_{t_f,t_i \to \infty}
\frac{W^{\mu\nu}_{t_f,t_i}(t, \vec Q) }{S_{t_f,t_i}} e^{- M_B t}\,, \nn \eea
where the numerator 
corresponds to  insertion of the three-dimensional Fourier
 transform of the (Euclidean) $T$-product  for $t>0$
\bea && W^{\mu\nu}_{t_f,t_i}(t, \vec Q) = \frac{1}{\pi} \int d^3 x \ e^{- i
\vec Q \cdot \vec x} \times \nn \\ &&
\langle 0 \vert   \Phi^\dagger_{\vec p_B=0}(t_f) J^\dagger_\mu(\vec x,
t) 
J_\nu (0)\Phi_{\vec p_B=0}(-t_i)\vert 0 \rangle \nn
 \label{eq:fun}
\eea
and the denominator is the usual $B$-meson propagator
\beq \label{eq:sb}  S_{t_f,t_i}=
\langle 0 \vert  \Phi^\dagger_{\vec p_B=0}(t_f)  \Phi_{\vec p_B=0}(-t_i)
\vert 0 \rangle 
\eeq
$\Phi_{\vec p_B}(t)$ is the $B$ interpolating
field with definite spatial momentum $\vec p_B$
\beq
\Phi_{\vec p_B}(t)=\int d^3x e^{-i\vec p_B\cdot\vec x}\Phi_B(\vec x,t)
\eeq
In terms of $f(k_+)$, we have 
\bea \label{eq:nclb} & \, & W^{\mu\nu}(t, \vec Q) = \\
&=&  \frac{1}{2\pi } \int_{0}^{M_B} dk_+ \  f(k_+) 
\frac{e^{- \left( k_+ +
\sqrt{ \vec Q^2} \right) t } }{2 \sqrt{ \vec Q^2}}
 \nn \\ && \times  \left( \bar Q^\mu \delta^{\nu 0}
+ \bar Q^\nu \delta^{\mu 0} - g^{\mu \nu} \bar  Q^0 - i \epsilon^{0 \mu \nu
\alpha } \bar Q_\alpha \right)   \nonumber \eea
where  
$Q^+_0 = k_+ +\sqrt{k_+^2 + \vec Q^2} \sim k_+ + \sqrt{ \vec Q^2} $.
By a suitable choice of the Lorentz components $\mu$ and $\nu$ of the currents
and of the spatial momentum $\vec Q$,
and by studying the time dependence of $W^{\mu\nu}(t, \vec Q)$,
 we can unfold  the integral above  and extract the shape 
function. As explained in~\cite{aglie1}, the same method 
can be used to extract the structure functions
of deep inelastic scattering.
\par {\bf Light-cone wave-funtions} The light-cone
wave functions  allow to predict  form factors 
relevant in many exclusive processes, such as 
  electromagnetic elastic scattering  at large momentum
transfer,  or    exclusive semi-leptonic decays  as $B \to \pi$ ($B \to \rho$) 
and   $B \to K^* \gamma$ decays~\cite{general}--\cite{ball}.
The knowledge of the relevant light-cone functions
would allow the determination of the form factors at $q^2 \sim 0$, a kinematical region
which is not accessible with standard lattice techniques.
In~\cite{aglie2}, it was shown that they can be computed, analogously
to the shape function, by suitable ratios of three- and two-point correlation
functions of Euclidean $T$-product. In this case the  three-point function
corresponds to the insertion of the three-dimensional Fourier transform
of the $T$-product
\be \langle \pi^- \vert T\Bigl[ \bar d(x)  \exp^{i \int_0^s A_\mu(s^\prime n^\mu) 
n^\mu ds^\prime} \gamma_\mu \gamma_5 u(0) \Bigr] \vert 0 \rangle  \ee
where $n^\mu=(q^\mu + p_\pi^\mu)/\sqrt{-q^2}$ and $x^\mu= s n^\mu$.
The OPE of the $T$-product can be written in terms of
 matrix elements of local operators
\be \label{eq:opsops} {\cal M}_n \propto 
\langle  \pi^- \vert \bar d(0)\gamma^\mu \gamma_5
(i D^{\mu_1}) \dots
(i D^{\mu_n}) u(0)  \vert 0 \rangle \ee
which, as for the shape function considered before,  
corresponds to the moments of the ligh-cone wave-fuction
\be \label{eq:phidef} {\cal M}_n = \int_0^1 du \ u^n \Phi_\pi(u)  \ee
The matrix elements of the 
local operators corresponding to the first two non-trivial moments
were studied, using the standard approach in~\cite{mms}.
With the OPEwOPE one can obtain the full light-cone wave function
$\Phi$ without the problems associated with the renormalization of the
local operators of eq.~(\ref{eq:opsops})~\cite{aglie2}. 
\section*{Acknowledgements}
I thank all my colleagues of the   APE  collaboration for 
very useful discussions. 
I am particularly indebted to V. Lubicz, C.T. Sachrajda and A. Vladikas.

\end{document}